\documentclass[prl,twocolumn,showpacs]{revtex4}

\usepackage{graphicx}
\usepackage{epstopdf}
\usepackage{xcolor}

\usepackage{amsmath}
\usepackage{amssymb}
\usepackage{graphicx}
\usepackage{dcolumn}
\usepackage{bm}

\newcommand{\del}{\Delta E^L(|p_i|)}
\newcommand{\der}{\Delta E^R(|p_i|)}
\newcommand{\detot}{\Delta E^T(|p_i|)}
\newcommand{\detl}{\Delta T^L(|p_i|)}
\newcommand{\detr}{\Delta T^R(|p_i|)}
\newcommand{\p}{|p_i|}

\begin{document}

\preprint{APS/123-QED}

\title{Matter, Energy, and Heat Transfer in a Classical Ballistic Atom Pump}

\author{Tommy A. Byrd$^{1}$, Kunal K. Das$^{2}$, Kevin A. Mitchell$^{3}$, Seth Aubin$^{1}$, and John B. Delos$^{1}$}

\affiliation{$^{1}$ Department of Physics, College of William and Mary, Williamsburg, VA 23187, USA }
\affiliation{$^{2}$ Department of Physical Sciences, Kutztown University of Pennsylvania, Kutztown, PA 19530, USA }
\affiliation{$^{3}$ School of Natural Sciences, University of California, Merced, CA 95344, USA}

\date{\today}

\begin{abstract}A ballistic atom pump is a system containing two reservoirs of neutral atoms or molecules and a junction connecting them containing a localized time-varying potential. Atoms move through the pump as independent particles. Under certain conditions, these pumps can create net transport of atoms from one reservoir to the other. While such systems are sometimes called ``quantum pumps,'' they are also models of classical chaotic transport, and their quantum behavior cannot be understood without study of the corresponding classical behavior. Here we examine classically such a pump's effect on energy and temperature in the reservoirs, in addition to net particle transport. We show that the changes in particle number, of energy in each reservoir, and of temperature in each reservoir vary in unexpected ways as the incident particle energy is varied.
\end{abstract}

\pacs{67.85.Hj, 05.60.Gg, 03.65.Sq, 37.10.Vz}
\maketitle
Particle transport is an ongoing topic of interest in a variety of systems including solid state electronics, microfluidic devices, and atomtronics components. In electronic solid-state systems, the transport of electrons through mesojunctions having time-dependent potential barriers, a phenomenon often called ``quantum pumping,'' has been theorized for decades \cite{thouless,brouwer-1,buttiker-floquet,das-opatrny,chamon-spin,GarttnerSchmelcher_PRE2010,Ferry-Goodnick}. It has been shown that such a system can pump electrons from one reservoir to another with no bias (such as a potential difference). More recently, Das and Aubin have proposed simulating such electron pumps using a system of neutral cold atoms with optical potentials as the driving forces \cite{DasAubin_PRL2009,Das2011,kunalpaddlewheel}. Neutral atom transport is becoming increasingly important in its own right due to the ongoing development of atomtronics, which seeks to replicate properties of electronics using neutral atoms, and in the field of quantum computing \cite{zhaochen}. Analogues of batteries, diodes, transistors, and recently hysteresis \cite{zoz,pepinocooper,eckel} have been explored in ultracold neutral atom systems.

Previous studies of these pumps have generally been within the quantum regime, and largely focus on charge or spin transport associated with fermionic carriers. In this paper we study the classical analogues of such pumps, and focus on the differential transfer of particles, energy, and heat. This broadens the study of quantum pumps into a new and largely unexplored regime. These classical analogues of quantum pumps are also interesting because they provide models of chaotic transport, which occurs in a great variety of systems on scales from nuclei to galaxies \cite{53,57,59,63,76,78,84,79,87,105,111}.

The pumps we consider are effectively one-dimensional, so the Hamiltonian is
\begin{eqnarray}
H(p,x,t) = p^2/2m + V(x,t).\label{ham}
\end{eqnarray}
We choose $V(x,t)$ to consist of two repulsive barriers, one or both of which oscillates. When both barriers oscillate they have the
same frequency $\omega$, but not the same phase. We examine the classical scattering of equal numbers of particles which approach such pumps from each reservoir with equal and fixed incident energy.

It has been shown that flows may be zero or negligible in pumps with idealized limits such as delta-function barriers or uniform phase-space density \cite{das-opatrny,double}. Here we show that under more realistic conditions, such pumps can generate significant net transfer of both matter and energy. Understanding heat flow is also essential for any transport mechanism, and is of fundamental importance for thermoelectric devices \cite{casati}. Studies that have been done in the context of mesoscopic pumps \cite{Buttiker-heat,Wang-heat} used a strictly quantum picture involving exchange of quasi-particles, and heat flow was shown to be outwards from the pump towards the reservoirs. The classical model discussed here is more appropriate for higher temperatures, and we show that the pump can heat or cool one or both reservoirs.

\textbf{Summary of results:} In previous papers \cite{double,ByrdDelos,single} we have shown that two-barrier pumps have the following properties when at least one barrier oscillates. (1)  These so-called ``quantum pumps'' provide nice models of classical chaotic scattering, and their behavior is governed by a heteroclinic tangle.  (2) Quantum theory shows that monoenergetic particles incident on periodically oscillating barriers have final energies equal to $E_n=E_i+n\hbar\omega$, where $E_i$ is their initial energy and $\omega$ is the frequency of the pump;  classical and semiclassical theories are needed to understand the range of $n$ and the heights of the peaks. (3)  Net pumping of particles from one reservoir to another can occur if monoenergetic particles approach the pump from both sides. (4) Pumping can go in either direction, depending on the incident energy and the pump parameters. (5)  The amount of pumping is very sensitive to incident energy and to pump parameters, and cannot be predicted without detailed calculation. (6) It is possible to design a ``particle diode'' which only allows net particle transport in one direction for low-energy incident particles, and in the opposite direction for high-energy incident particles.

In this paper we show that for monoenergetic incident particles: (A) Such pumps can transfer energy from one reservoir to the other, and energy can be transferred from pump to particles or vice versa. (B) A net change of energy in each reservoir can occur even if there is no net particle transport. The direction of energy change is distinct from the direction of particle transport. (C) Such pumps can heat or cool one or both reservoirs, and the heating or cooling is distinct from the existence or direction of net particle transport and distinct from energy flow. (D) At some incident energies, such pumps can generate net particle transport while at the same time particles give energy to the pump.

\textbf{System} We will establish properties (A)-(D) by examining one specific pump: a particle diode consisting of two Gaussian-shaped potential barriers, only one of which oscillates. We choose this pump as our example because the dynamics of the system become much more complicated when both barriers oscillate \cite{double}. However, allowing the second barrier to oscillate (or changing the barrier parameters) only affects the conclusions discussed below quantitatively. Therefore properties (A)-(D) apply to general ballistic atom pumps.

In the chosen diode, the distance between the barriers is substantially larger than their widths, so their overlap is negligible.  The right-hand barrier has a fixed height, while the left-hand barrier oscillates between zero and the height of the right-hand barrier.  The pump is described by

\begin{align}
V(x,t)&=\hat{U}_L \left(1+\alpha_L \cos(\omega t)\right)\exp\left(\frac{-(x+\hat{x})^2}{2\sigma^2}\right) \label{gaussleft}\nonumber \\
&+\hat{U}_R \exp\left(\frac{-(x-\hat{x})^2}{2\sigma^2}\right),
\end{align}
where $\hat{U}_{L,R}$  is the average height of each barrier, $\alpha_{L}$ is the amplitude of
oscillation of the left barrier, $\omega=2\pi/T$ is the frequency and $T$ is the period, and $\sigma$ is the standard deviation of each Gaussian. The left and right barriers are centered at $x=-\hat{x}$ and $x=\hat{x}=4.5$, respectively. In our calculations, we set $\hat{U}_L=1$, $\hat{U}_R=2\hat{U}_L=2$, $\alpha_L=1$, $\omega=1$, $\sigma=1.5$, and $m=1$. These are scaled units \cite{single}.

The effects of the pump can be understood qualitatively as follows. For incident energies less than the height of the static barrier, all particles from the right reflect from the static barrier, but particles incident from the left may gain enough energy from the oscillating barrier to scatter past both barriers. Consequently, the only possible direction of net particle transport is from left-to-right. For incident energies greater than the height of the static barrier, computations show that, in this case, all particles incident from the right transmit past both barriers, but particles incident from the left may lose energy to the oscillating barrier, reflect from the static barrier, and ultimately scatter to the left reservoir. Thus the only possible direction of net particle transport reverses to right-to-left.

\textbf{Method}
Our computational algorithm can be summarized as follows: 1) For each
initial energy, launch particles toward the barriers from the left and right. Particles begin with a range of positions $\Delta x=|p_i|2\pi/\omega$  where $p_i$ is the initial momentum, which ensures that all barrier phases are encountered. 2) Record the reservoir to which each particle is scattered, and sum the results to obtain the net fractional transport (defined below) of particles scattered to the right (which may be negative if more particles are scattered to the left).
3) Compute the total energy gain of the two reservoirs after scattering, which may be negative if the system loses energy to the pump. 4) Compute the net gain (or loss) in the total energy of each reservoir. Energy being an extensive quantity, a reservoir gains total energy by gain in the number of particles as well as by gain of energy of individual particles passing though the pump.
5) Compute the change of energy of each particle scattered into each reservoir, and compute the average of these changes for all particles scattered into each reservoir. The average change of energy per scattered particle may be regarded as corresponding to a change of temperature of the reservoir.  Temperature being an intensive property, the direction of temperature change need not be the same as the direction of energy change in each reservoir. Formulas for computation of these quantities are given below.

The fractional transport of particles through the pump is defined as
\begin{align}
C_P(|p_i|)=\frac{R(|p_i|)-L(|p_i|)}{R(|p_i|)+L(|p_i|)},\label{partcurr}
\end{align}
where $R(|p_i|)$ is the number of particles scattered to the right for each $|p_i|$, and $L(|p_i|)$ is the number of particles scattered to the left. The sum $R(|p_i|)+L(|p_i|)$ represents all particles incident on the pump for a given $|p_i|$. $C_P(|p_i|)$ is positive when more particles are scattered to the right (net particle transport to the right reservoir), negative when more particles are scattered to the left (net particle transport to the left reservoir) and zero when equal numbers of particles scatter to the right and left reservoirs.

\begin{figure*}[t]
\includegraphics*[width=\textwidth]{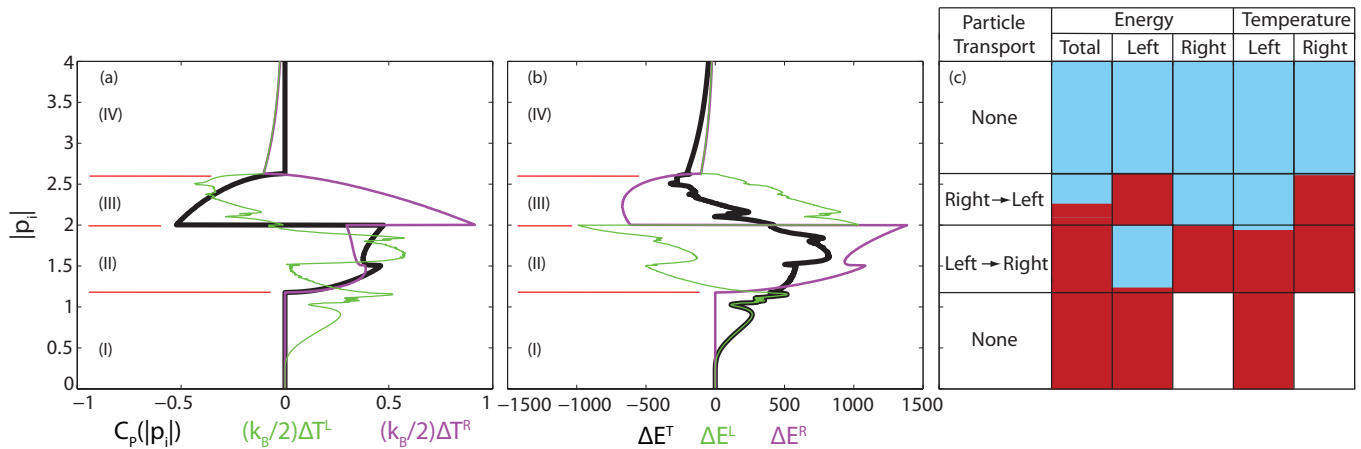}
\caption{(Color online) (a) Net particle transport, $C_P(|p_i|)$ [thickest curve], and the change in average energy per particle in the left reservoir, $(k_B/2)\detl$ [thin (green) curve] and right reservoir, $(k_B/2)\detr$ [medium (purple) curve]. When $C_P(|p_i|)$ is positive (negative) there is net particle transport from left-to-right (right-to-left). When $(k_B/2)\Delta T^{L,R}(\p)$ is positive (negative), the pump increases (decreases) the average energy of particles scattered into the respective reservoir, and the temperature in that reservoir increases (decreases). (b) Total energy change in both reservoirs, $\detot$ [thickest curve], in the left reservoir, $\del$ [thin (green) curve], and in the right reservoir, $\der$ [medium (purple) curve]. When $\detot$ is positive, the pumps adds net energy to the reservoirs; when negative, the reservoirs lose net energy to the pump. When $\Delta E^{L,R}(\p)$ is positive (negative), the pump increases (decreases) the total energy in the respective reservoir. (c) Summary of (a) and (b). Dark gray (red) indicates an increase, and light gray (blue) represents a decrease. No color is plotted if the quantity does not change.}\label{pic}
\end{figure*}

For each initial particle energy, the total energy change of the system and each reservoir are defined as
\begin{equation}
\Delta E^\alpha(|p_i|) = E_f^\alpha(|p_i|)-E_i^\alpha(|p_i|),
\end{equation}
where $\alpha=\{T,L,R\}$. When $\alpha=T$, $E_f^T(|p_i|)$ and $E_i^T(|p_i|)$ represent the total final and initial energies, respectively, of all particles incident upon one cycle of the pump. When $\Delta E^T>0$, the pump has added energy to the reservoirs; when $\Delta E^T<0$, the reservoirs have lost energy to the pump. When $\alpha=L$ or $R$, $E_f^{\alpha}(|p_i|)$ represents the total final energy of all particles which scatter to the left or right reservoirs, and $E_i^{\alpha}(|p_i|)$ represents the corresponding total initial energy of all particles beginning in the left or right reservoir.

The last quantities examined in this paper are the changes in \emph{average} energy per particle scattered into each reservoir. These quantities are defined as
\begin{align}
\overline{\Delta E^\beta(\p)}=\frac{E_f^\beta(|p_i|)}{M^\beta}-\frac{E_i^\beta(|p_i|)}{N^\beta}
=\frac{k_B}{2}{\Delta T^\beta(|p_i|)},\label{temp}
\end{align}
where $\beta=\{L,R\}$ and corresponds to the left and right reservoirs, respectively. $N^\beta$ is the number of particles incident on the pump from the $\beta$ reservoir in one cycle, and $M^\beta$ is the number of particles scattered to the $\beta$ reservoir. A total of $2N^\beta$ particles approach the pump for each incident energy ($N^\beta$ from each reservoir); consequently $M^\beta>N^\beta$ corresponds to an increase in particle number for the $\beta$ reservoir. This change of average energy per particle can be regarded as a change of temperature of those scattered particles. Then a positive (negative) $\Delta T^{L,R}$ produces an increase (decrease) in the temperature of the corresponding reservoir after thermalization.

\textbf{Results} In Fig.~\ref{pic} we show the results of calculations for net particle transport, energy changes in the total system, and temperature and energy changes in each reservoir for the selected pump. We discuss all properties in relation to the net particle transport, which is the thick curve in Fig.~\ref{pic}(a). There are four distinct regions of particle transport direction, and we discuss them in order of increasing complexity. This complexity arises for two reasons.  (1) Depending on the initial energy and the frequency of the barrier, a particle can ride repeatedly up and down the oscillating barrier.  (2)  A particle can undergo multiple reflections between the two barriers;  this is the source of chaos in the system.

\textbf{Region I: No particle transport; left reservoir heated} ($0<|p_i|\lesssim 1.176$) At these low energies, no particle gets past the static barrier, so there is no net particle transport, and $C_P(|p_i|)=0$ [thickest curve in Fig.~\ref{pic}(a)]. Particles incident from the right reflect from the static barrier into the right reservoir without a change in energy. Therefore the number of particles, their average energy, and the total energy in the right reservoir do not change, i.e. $\detr=0$ [medium curve (purple online) in Fig.~\ref{pic}(a)] and $\Delta E^R(|p_i|)=0$ [medium (purple online) curve in Fig.~\ref{pic}(b)].

All particles incident from the left are scattered into the left reservoir, but the oscillating barrier changes their energy. They may gain or lose energy to the pump, depending on their time of arrival. On average, they gain energy. Accordingly, the temperature (average energy per particle) and total energy both rise in the left reservoir, i.e., $\detl>0$ and $\del>0$ [thin (green online) curves in Fig.~\ref{pic}(a) and~\ref{pic}(b)]. Considering both reservoirs together, there has been net addition of energy from the pump to the reservoirs ($\detot>0$) [thickest curve in Fig.~\ref{pic}(b)], and this energy is entirely added to the left reservoir.

These results are summarized in Fig~\ref{pic}(c), in which the light gray (blue online) represents a loss, dark gray (red online) represents an increase, and white represents no change.

\textbf{Region IV: No particle transport; both reservoirs cooled} ($\p\gtrsim2.63$) At high incident momentum, all particles incident from both sides transmit past both barriers, and there is no net particle transport ($C_P(\p)=0$). Particles incident from both sides lose energy (on average) to the pump, which causes a decrease in the total energy of each reservoir ($\Delta E^{L,R}(\p)<0$) and total energy of the system ($\detot<0$). The average energy changes of particles scattered into each reservoir are equal ($\detl=\detr<0$) and each reservoir is cooled. Fig.~\ref{pic}(c) summarizes these results. Calculations show that the loss of energy to the pump decreases exponentially with $|p_i|$, a result that calls for a general proof.

\textbf{Region II: Net left-to-right particle transport} ($1.176\lesssim |p_i| \leq 2$) This region is defined by the fact that all particles from the right are reflected by the static barrier, but some particles incident from the left gain enough energy from the pump to scatter into the right reservoir. Accordingly, the right-hand reservoir gains particles ($C_P(|p_i|)>0$) and average evergy per particle ($\detr>0$), and the reservoir is heated. The total energy of the reservoir increases ($\der>0$).

Some particles which begin on the left scatter to the left, and the pump can change their energy. Over most of region II ($1.176\lesssim|p_i|\lesssim 1.95$), the left-to-left scatterers gain energy from the pump (on average) ( $\detl>0$), and temperature of the left reservoir increases. However at the high end of this region ($1.95 \lesssim \p <2$), the left-to-left scatterers lose energy (on average) to the pump ( $\detl<0$), and the left reservoir is cooled.

The total energy change of this reservoir depends on the average energy change of left-scattered particles, and on the loss of particles to the right-hand reservoir.  Over most of region II ($1.243\lesssim \p <2$), there is a net loss of energy in the left-hand reservoir ($\del<0$). However at the lower end of this region ($1.176\lesssim|p_i|\lesssim 1.243$), the gain of energy of left-to-left scatterers exceeds the loss of energy associated with particle transport to the right, and the total energy in the left-hand reservoir rises ($\del>0$)

Combining the energy changes of both reservoirs, the pump has added energy to the reservoirs for the entirety of region II ($\detot>0$). These results are summarized in Fig~\ref{pic}(c).

\textbf{Region III: Net right-to-left particle transport} ($2<|p_i|\lesssim 2.63$) This region is the most complex. For $|p_i|>2$, for this pump, all particles incident from the right have enough energy to transmit past both barriers. Particles incident from the left initially have enough energy to get over the static barrier, but they may lose energy to the oscillating barrier, be reflected from the static barrier, and scatter into the left reservoir.  Therefore the only possible direction of net particle transport is from right-to-left. Fig.~\ref{pic}(a) shows right-to-left particle transport ($C_P(|p_i|)<0$) in the range $2<|p_i|\lesssim 2.63$.

Particles which scatter to the right reservoir begin in the left reservoir. In the majority of this region ($2<|p_i|\lesssim 2.616$) they (on average) gain energy from the pump ($\detr>0$), and the temperature in the right reservoir rises. Combining the gain of energy per particle with the loss of particles, the result is a loss of total energy in the right reservoir ($\der<0$). In the remainder of region III, ($2.616<|p_i|\lesssim 2.63$), the left-to-right scatterers lose energy to the pump (on average) ($\detr<0$), the right reservoir is cooled, and its total energy decreases ($\der<0$) because of loss of particles and loss of average particle energy.

Particles which scatter to the left reservoir can begin in either reservoir. These particles on average lose energy to the pump ($\detl<0$), so the left reservoir is cooled. However its total energy rises ($\del>0$) because scattering increases particle number in the reservoir. Examining both reservoirs together, over most of the lower portion of region III ($2<|p_i|\lesssim 2.267$), the pump adds energy to the reservoirs, while over the remainder of the region ($2.267\lesssim|p_i|\lesssim 2.63$), it removes energy from the reservoirs. Fig.~\ref{pic}(c) summarizes these results.

\textbf{Averaging over energies} Thermodynamics (and physical intuition) tell us that if a pump is connected to a single reservoir (or two reservoirs with the same temperature, pressure, and chemical potential) then the net energy transfer can only go from the pump to the reservoirs. Accordingly if we average the energy input $\detot$ over a Maxwellian distribution at any temperature, that result must be nonnegative ($\int{\detot e^{-p_i^2/2mk_BT}dp}\geq 0$). Scrutiny of $\detot$ in Fig.~\ref{pic}(b) shows that this is satisfied in the example pump. Also the observation that at low incident particle energies (Region I), the net energy flow is from pump to particles must hold for any pump. This is another point that calls for a dynamical proof.

We have therefore by example established the properties (A)-(D) stated in the ``Summary of results.''

\textbf{Acknowledgements} KKD acknowledges support of the NSF under Grant No. PHY-1313871, and of a PASSHE
-FPDC grant and a research grant from Kutztown University..  KAM acknowledges NSF support from Grant No. PHY-0748828. JBD and TAB acknowledge NSF support via Grant No. PHY-1068344. Calculations were performed on the SciClone computing complex at The College of William and Mary, which were provided with the assistance of the National Science Foundation, the Virginia Port Authority, Sun Microsystems, and Virginia's Commonwealth Technology Research Fund.


\end{document}